\documentclass{appolb}
\usepackage{graphicx}
\usepackage{amssymb}

\newcommand{\ncfg}{N_{\mbox{cfg}}}
\newcommand{\pslash}{\not\!p}
\newcommand{\Kslash}{\not\!K}

\begin{document}

\title{Lattice Landau gauge quark propagator and the quark-gluon vertex
\thanks{Presented by O. Oliveira at Excited QCD 2016, Portugal,  2016}}

\author{Orlando Oliveira$^1$, Ay{\c s}e K{\i}z{\i}lers\"u$^2$, Paulo J. Silva$^1$, Jon-Ivar Skullerud$^{2,3,4}$, Andre Sternbeck$^5$, Anthony G. Williams$^{2,6}$
\address{$^1$CFisUC, Department of Physics, University of Coimbra, P--3004 516 Coimbra, Portugal \\ 
$^2$ CSSM, Department of Physics, University of Adelaide, Adelaide, SA 5005, Australia\\
$^3$Department of Mathematical Physics, National University of Ireland Maynooth, Maynooth, County Kildare, Ireland \\ 
$^4$School of   Mathematics, Trinity College, Dublin 2, Ireland \\
$^5$Theoretisch-Physikalisches Institut, Friedrich-Schiller-Universit\"at Jena, 07743 Jena, Germany \\
$^6$ ARC Center of Excellence for Particle Physics at the Terascale, Department of Physics, University of Adelaide, Adelaide, SA 5005, Australia }
}
\maketitle
\begin{abstract}
We report preliminary results of our ongoing lattice computation of the Landau gauge quark propagator and the 
soft gluon limit of the quark-gluon vertex with 2 flavors of dynamical $\mathcal{O} (a)$ improved Wilson fermions.
\end{abstract}
%
  
\section{Introduction}

The dynamics of a quantum field theory is encoded in its Green's functions.
Here, we focus on two fundamental correlation functions of QCD, namely the quark propagator and the 
quark-gluon vertex, in the Landau gauge.

The quark propagator has two form factors, the quark wave function and the quark running mass.
The running mass signals the dynamical symmetry breaking mechanism present in QCD. 
Furthermore, from the quark propagator one can access its
analytical properties which, hopefully, provide information on quark confinement. 

A complete description of the quark-gluon vertex requires the computation of twelve form factors. This three point correlation
function is of primary importance for hadronic physics and encodes information on the hadronic spectrum and on quark confinement.

Lattice simulations allow for first principles calculations of both the quark propagator and the quark-gluon
vertex form factors, or combinations of form factors, associated with the above
mentioned correlation functions. Once the finite size effects are under control, the results from lattice QCD simulations
may also be used in the validation of continuum approaches to strong interaction physics as those provided by Dyson--Schwinger 
or functional renormalisation group equations.

The quark propagator and guark-gluon vertex are gauge dependent quantities. Here, we report on results for the Landau gauge 
defined, in the continuum, by the condition $\partial_\mu A^a_\mu = 0$, 
where $A^a_\mu$ refers to the gluon field. On the lattice, the Landau gauge condition is satisfied up to $\mathcal{O}(a^2)$ corrections
in the lattice spacing.

The gauge ensembles in our study use three values for the gauge coupling, corresponding to lattice spacings of 
$a\approx0.081$ fm, $a\approx0.071$ fm and $a\approx0.060$ fm, and we use quark masses corresponding to 
$m_\pi\approx290$ MeV and $m_\pi\approx420$ MeV. The ensembles were generated by the 
Regensburg QCD (RQCD) collaboration~\cite{Bali:2014} 
with $N_f=2$ non-perturbative improved Sheikholeslami--Wohlert (clover) fermions\footnote{We used the ensembles I, III, IV, VI and X
referred to in~\cite{Bali:2014} and thank the RQCD collaboration for providing us access to them.}. 
The gauge fixing and the computation of the propagators and vertex functions were performed on
the HLRN~\cite{HLRN} supercomputing facilities.
The lattice setup and the parameters used are listed in table~\ref{tab:params}.

\section{The Quark Propagator}

\begin{table}[t]
\begin{center}
\begin{tabular}{cc|ccccc}
$\beta$ & $\kappa$ & $a$ [fm] & $V$ & $m_\pi$ [MeV] & 
 $m_q$ [MeV] &  $\ncfg$ \\ \hline
5.20 & 0.13596 & 0.081 & $32^3\times64$ & 280 & 
  6.2 & 900 \\
5.29 & 0.13620 & 0.071 & $32^3\times64$ & 422 & 
 17.0 & 900 \\
5.29 & 0.13632 & 0.071 & $32^3\times64$ & 295 & 8.0      & 908 \\
     &         &       & $64^3\times64$ & 290 &          & 314 \\
5.40 & 0.13647 & 0.060 & $32^3\times64$ & 426 & 
 18.4 & 900
\end{tabular}
\end{center}
\caption{Lattice setup: the lattice spacing and pion masses are taken from~\cite{Bali:2014}.
              The subtracted bare quark mass is given by $m_q=1/(2\kappa)-1/(2\kappa_c)$ and
              we read the critical hopping parameters $\kappa_c$ from~\cite{Bali:2014}.}
\label{tab:params}
\end{table}

The continuum momentum space quark propagator is given by
\begin{equation}
S(p) = \frac{1}{i\pslash A(p^2) + B(p^2)} =
\frac{Z(p^2)}{i\pslash+M(p^2)} \ ,
\end{equation}
where $Z(p^2) = 1/A(p^2)$ is the quark wave function and $M(p^2)= B(p^2)/A(p^2)$ is the renormalisation group invariant running mass function.
Our lattice simulations use periodic boundary conditions in space and anti-periodic boundary conditions in time for fermion fields 
and, therefore, the available momenta read
\begin{eqnarray}
p_i  & = &\frac{2\pi}{N_i a}\left(n_i - \frac{N_i}{2}\right) ;\qquad n_i =  1,2,\cdots,N_i\,, \\
p_t & = & \frac{2\pi}{N_t a} \left(n_t-\half-\frac{N_t}{2}\right) ;\qquad  n_t =  1,2,\cdots,N_t \,,
\label{eq:latt_momenta}
\end{eqnarray}
where $N_i, N_t$ refers to the number of lattice points in the spatial and
temporal directions, respectively.  For Wilson fermions, the lattice quark propagator is given by
\begin{equation}
S^L(pa) = \frac{Z^L(pa)}{ia\Kslash(p)+aM^L(pa)}\,,
\label{eq:prop-lat}
\end{equation}
where the lattice momentum variable is
\begin{equation}
K_\mu(p) = \frac{1}{a}\sin(ap_\mu)\,.
\end{equation}

\begin{figure}[t]
\centerline{
\includegraphics*[width=6cm]{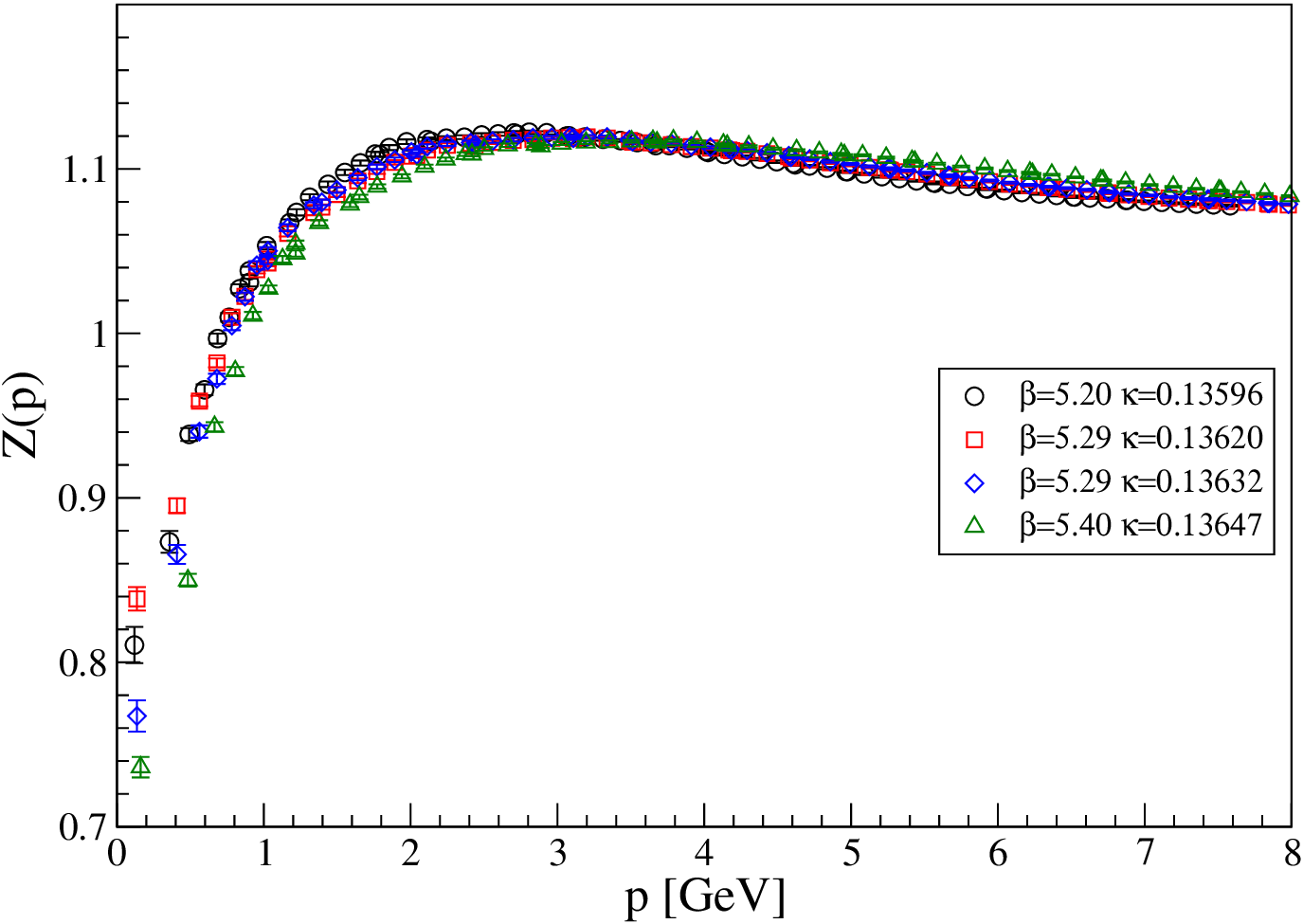} ~
\includegraphics*[width=6cm]{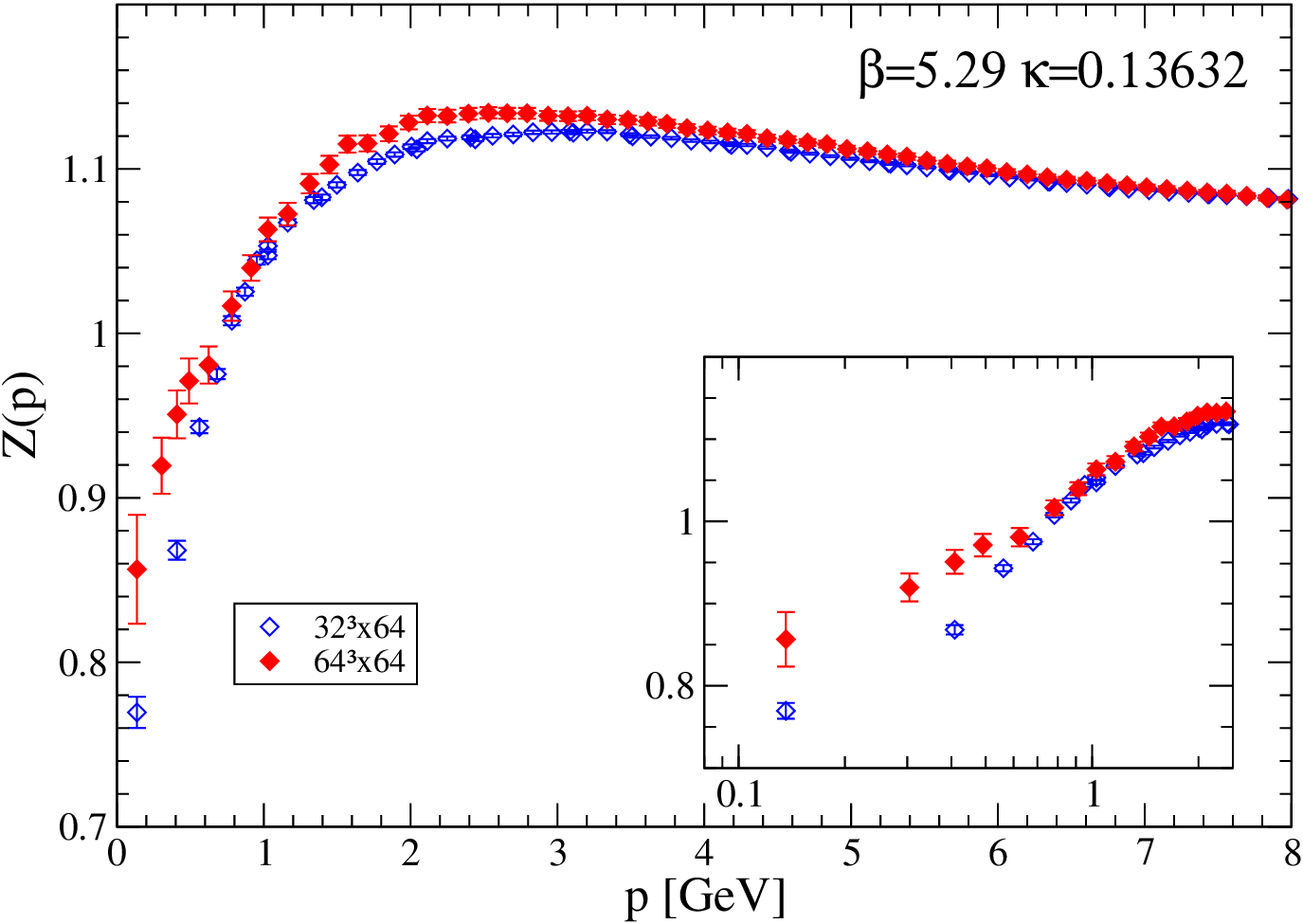}}
\caption{Tree-level corrected wave function $Z(p)$ for the $32^3\times 64$ ensembles (left) and 
              for the two lattice volumes at $\beta=5.29, \kappa=0.13632$ (right),  as a function of the 
              momentum $p$. The momenta have been cylinder cut with a radius of 1 lattice momentum unit.}
\label{fig:Z-all}
\end{figure}

\begin{figure}[t]
\centerline{
\includegraphics*[width=6cm]{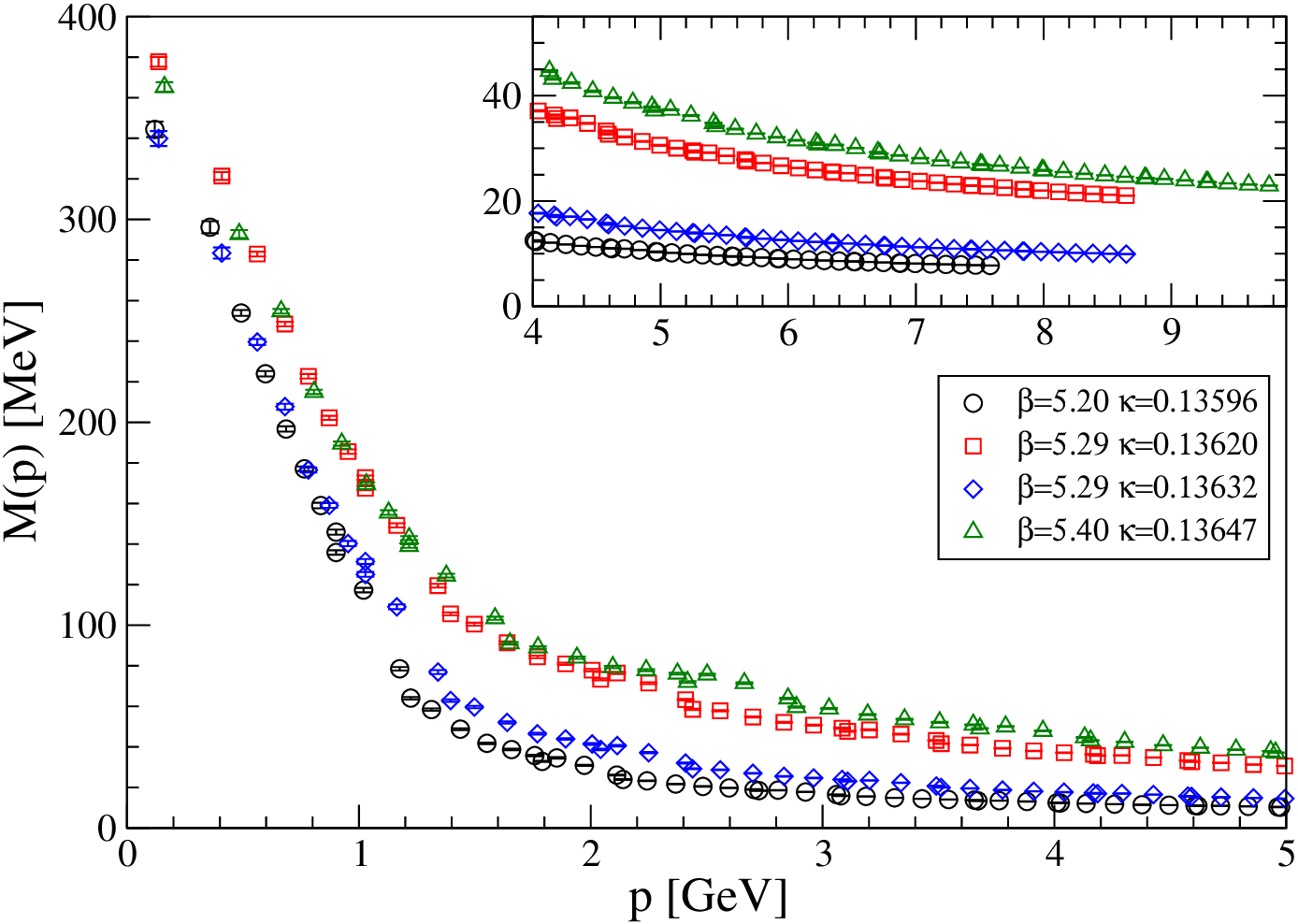} ~
\includegraphics*[width=6cm]{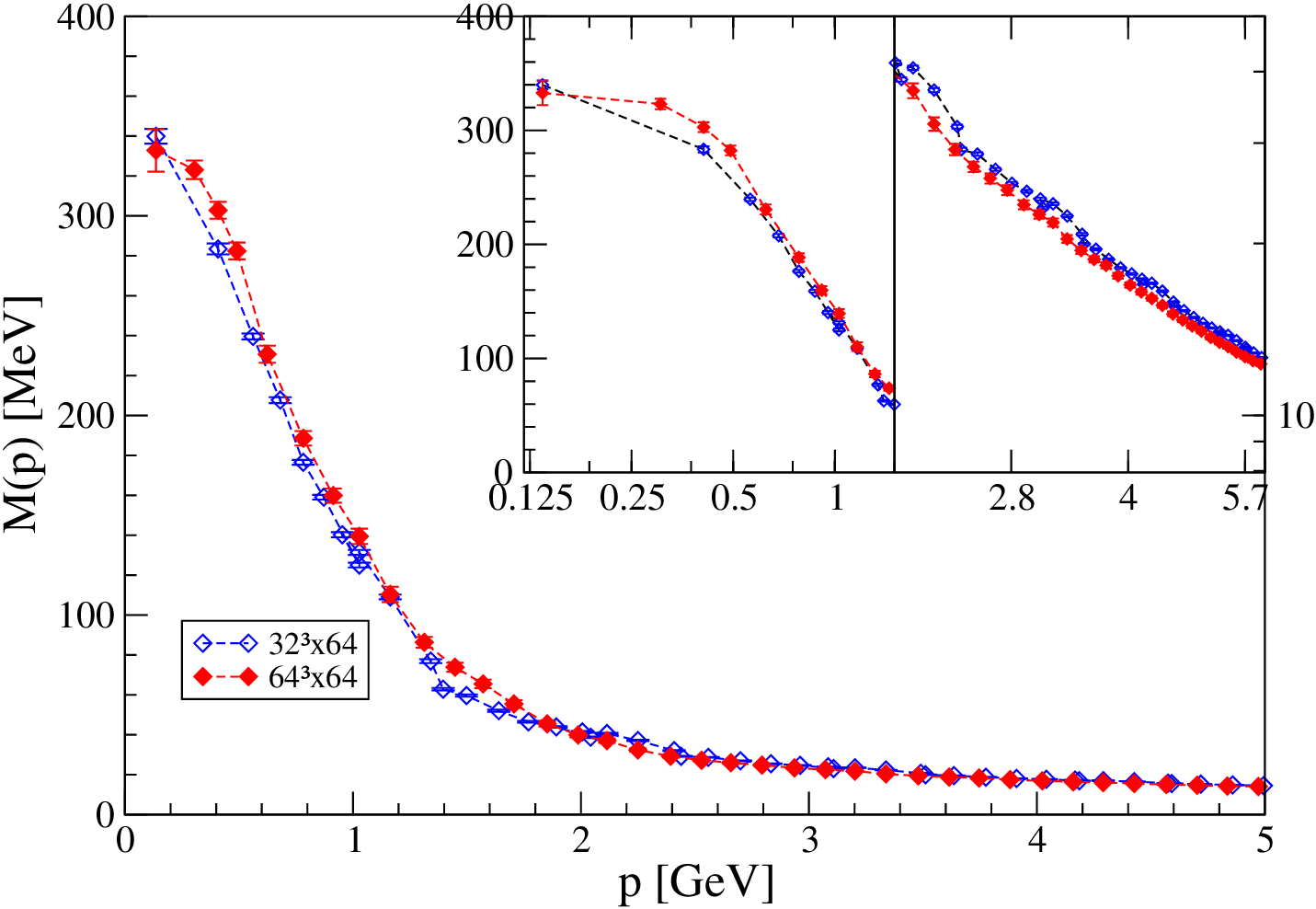}}
\caption{Tree-level corrected running mass $M(p)$ for the $32^3\times 64$  ensembles (left) and  
              for the two  lattice volumes at $\beta=5.29, \kappa=0.13632$ (right), as a function of the 
              momentum $p$.  On the left plot, the inset  a close-up of the high-momentum region.
             On the right plot, the two insets show a close-up of the low-momentum region with a 
             logarithmic momentum scale (left) and the high-momentum region on a
             log-log scale (right). The momenta have been cylinder cut with a radius of 1 lattice momentum unit.}
\label{fig:M-all}
\end{figure}

On the lattice, given that the rotational symmetry is  explicitly broken and one has to use a finite lattice spacing, 
large lattice artefacts are introduced in the quark propagator form factors. Here, we rely on tree level lattice
perturbation theory estimation of the rotated propagator, in combination
 with momentum cuts, to correct for these artefacts -- see~\cite{Skullerud2001a,Skullerud2001b}
for details. The results reported here are derived using 
the so called hybrid correction scheme which provide smoother form factors\footnote{The multiplicative
correction scheme introduces some artificial structures on the mass function at medium momenta. The running masses computed using
both the correction schemes define mass functions which are, at the qualitative level, equal with the multiplicative scheme producing larger
mass values. The two schemes seem to converge as we go towards higher momentum.}. We will also show the results for the so called H4 estimations~\cite{Becirevic1999,Soto2007} for the correction of the hypercubic artefacts.
The tree-level corrected wave function and running mass for the various ensembles can be seen in Figs.~\ref{fig:Z-all} and~\ref{fig:M-all}.

\begin{figure}[t]
\centerline{
\includegraphics*[width=6cm]{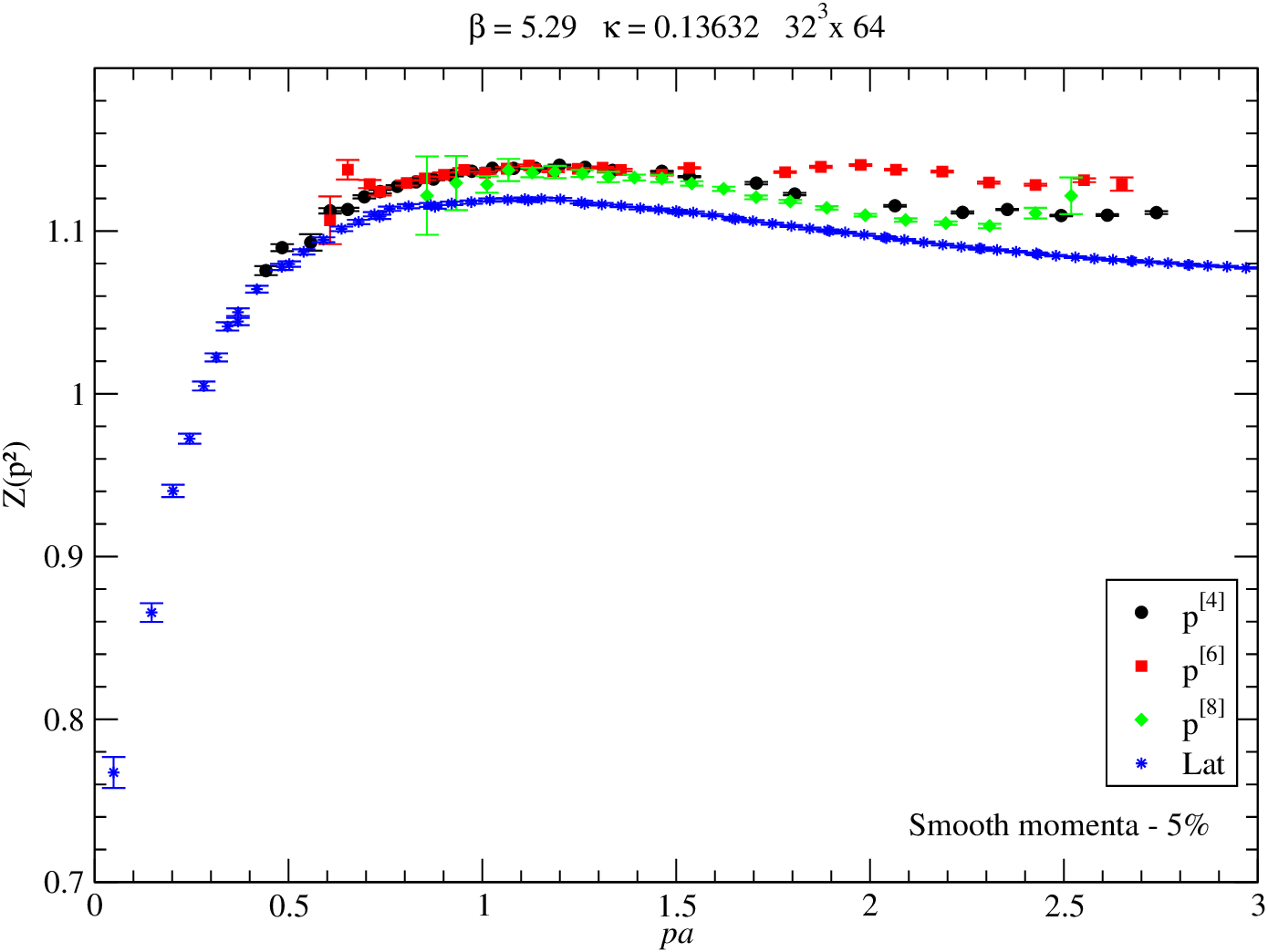} ~
\includegraphics*[width=6cm]{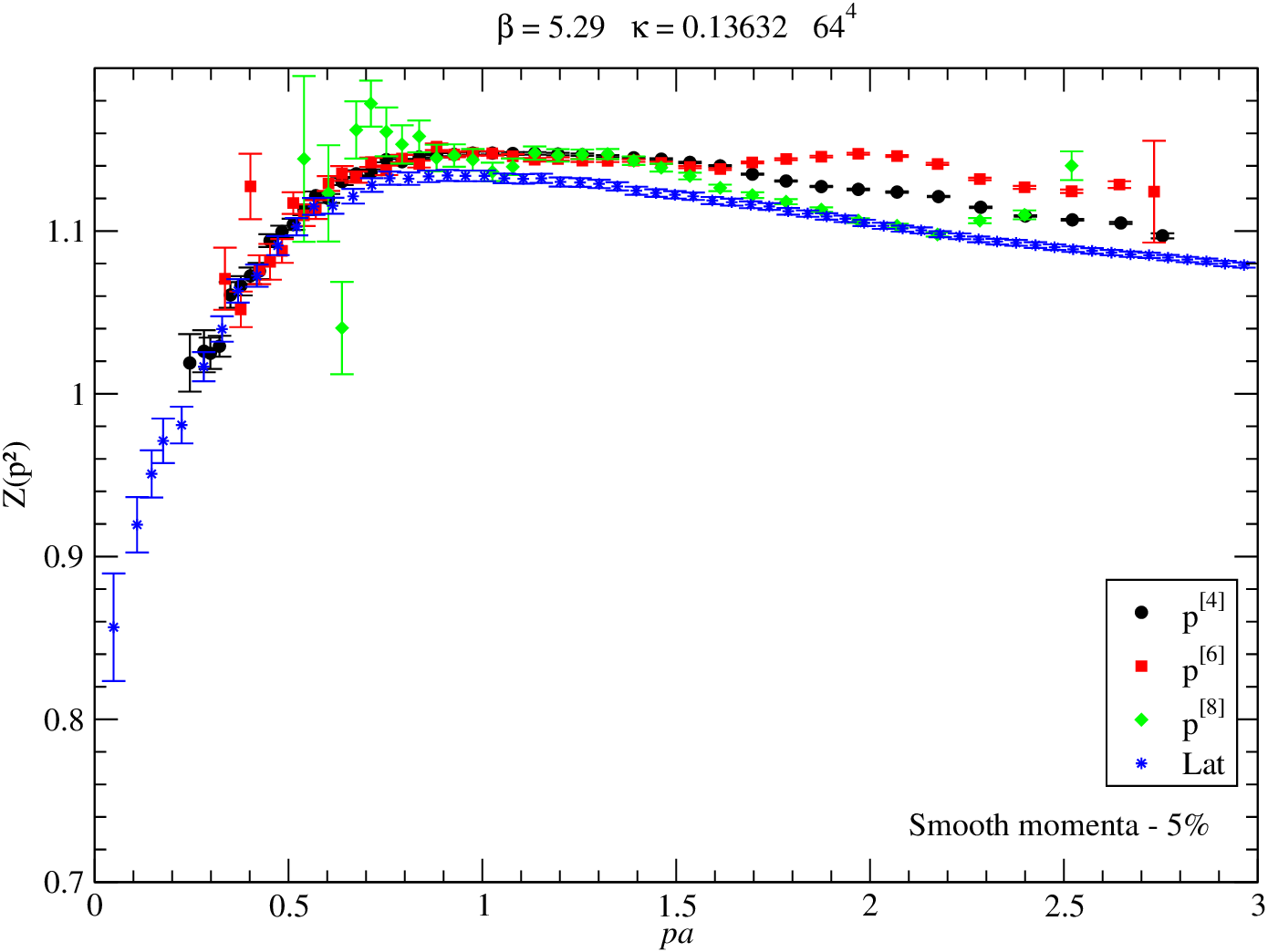}}
\caption{H4 extrapolation for the quark wave function $Z(p)$ for the $32^3\times 64$ ensemble (left) and 
              and $64^4$ ensemble (right) at $\beta=5.29$ and  for $\kappa=0.13632$.}
\label{fig:Z-ext}
\end{figure}

The quark wave function $Z(p^2)$ is suppressed in the infrared region and is a slightly decreasing function of the momentum for
$p \gtrsim 3$ GeV. For sufficiently large momentum $Z(p^2)$ should reproduce the perturbative result which predicts a constant
wave function in the Landau gauge. The observed slow decrease of $Z(p^2)$ is an indication that the subtraction scheme 
does not completely remove the lattice artefacts. 
In Fig.~\ref{fig:Z-ext} we rely on the H4 method to extrapolate away rotational symmetry violating lattice artefacts
for the quark wave function. In the range of momenta where the various extrapolations are compatible, the H4 method 
yields larger values for $Z(p^2)$ compared to the tree level corrections and predicts an essentially constant wave function for 
$pa$ above $\sim 0.8$.
The results of Fig.~\ref{fig:Z-all} suggest that $Z(p^2)$ has only a mild dependence on the quark mass at large momentum.
In contrast at low momentum the data also show an infrared suppression that becomes larger for the ensembles with
smaller bare quark masses.  Figure~\ref{fig:Z-all} also reveals clear finite volume effects in the infrared region and mild volume effects
at higher momenta and up to  $p \lesssim 3$ GeV.

The running quark mass, see Fig.~\ref{fig:M-all}, is a decreasing function of momentum and, in the infrared, $M(p^2)$ decreases
as $m_\pi$ approaches its physical value. For  smaller pion masses our simulations give $M(0) \approx 320$ MeV. 
The plots in Fig.~\ref{fig:M-all} show finite size effects in $M(p^2)$ which are not removed by the H4 method. Indeed, the
H4 method is not able to accommodate well the dependence of the running mass with the various momentum invariants.

\section{The quark-gluon vertex}

In general, the quark-gluon vertex is described by twelve form factors which are functions of $p^2$, $q^2$, $k^2$, respectively, 
the momentum of the outgoing quark, the gluon momentum and the momentum of the incoming quark. However, for the soft gluon limit,
where the gluon momentum vanishes, the vertex requires only three form factors \cite{Skullerud2003} and is given by
\begin{eqnarray}
\Gamma_\mu (p^2) ~~ = ~~
 \lambda_1 (p^2) \, \gamma_\mu ~ - ~ 
                   \lambda_2 (p^2) \, 4  \, p \!\!\! / \, p _\mu  ~ - ~
                   \lambda_3 (p^2) \, 2 \, i \,p_\mu \ .
\end{eqnarray}

\begin{figure}[t]
\centerline{
\includegraphics*[width=9cm]{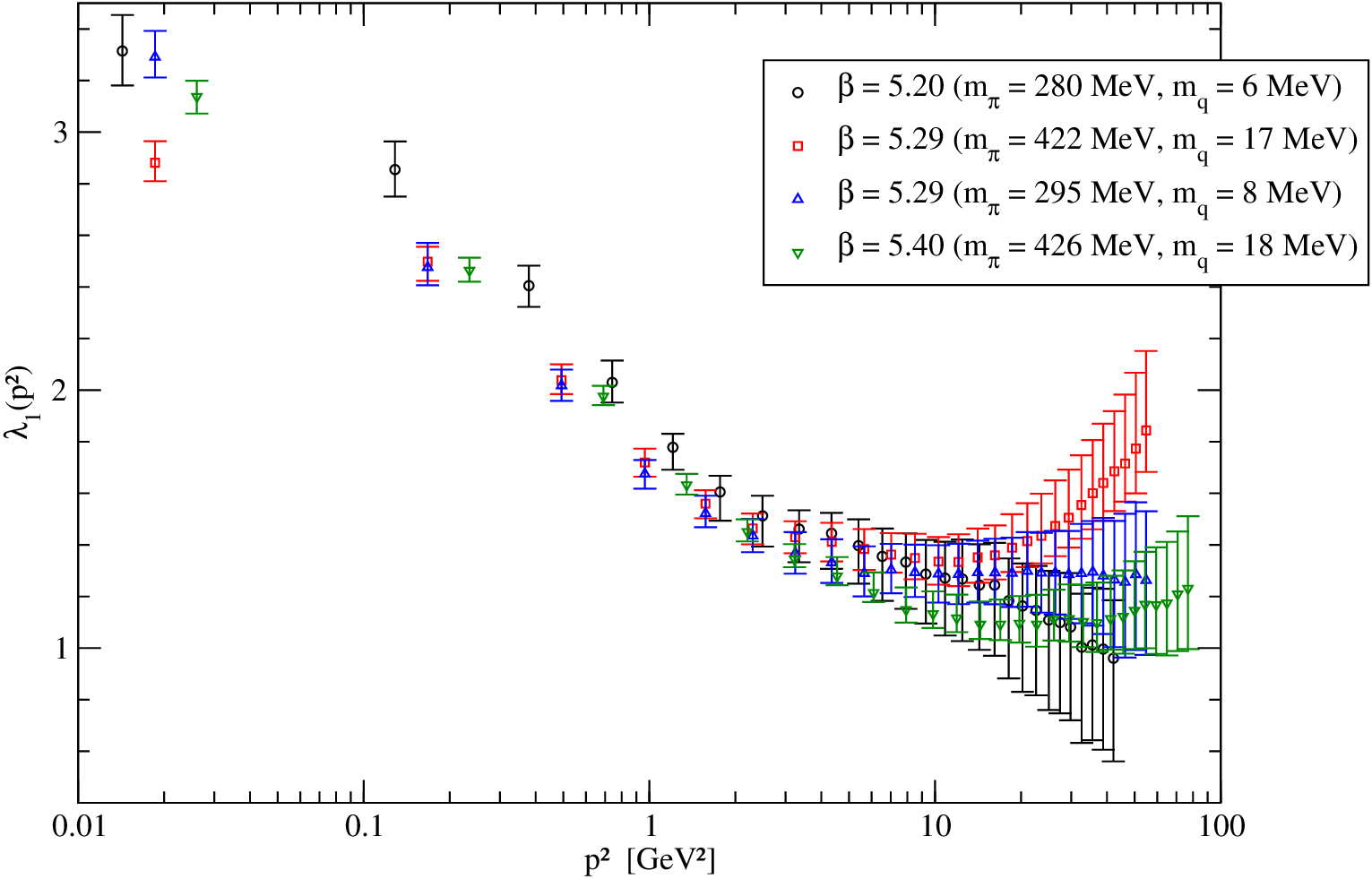}}
\caption{Tree-level corrected $\lambda_1 (p^2)$ from the $32^3\times 64$ ensembles.}
\label{fig:l1}
\end{figure}

In Fig.~\ref{fig:l1} 
we report on the lattice estimates for $\lambda_1$ after performing the tree level
corrections for the asymmetric ensembles referred in Tab.~\ref{tab:params}. 

The form factor $\lambda_1 (p^2)$, which can be used to define the strong coupling constant, is enhanced at infrared momenta for
all simulations. Due to the finite volume and lattice spacing effects, the dependence of $\lambda_1 (p^2)$ on the pion
mass is not clear from the plot. A better understanding of the finite size effects is required before drawing further conclusions.

\section{Summary and Outlook}

In this proceeding we report preliminary results for the quark propagator and quark-gluon vertex in the Landau gauge using lattice QCD
simulations. We are currently working towards providing results closer to the physical pion mass
and are trying to understand the effects of the lattice artefacts on both the wave function and running mass. On the other hand, for the
vertex form factors, we are engaged in computing the three form factors that describe the vertex in the soft gluon limit for all the ensembles 
in Tab.~\ref{tab:params}. We hope to provide final results in a not so distant future.

\end{document}